\documentclass[journal]{IEEEtran}
\usepackage{amsmath}
\usepackage{graphicx}
\usepackage{siunitx}
\usepackage{xcolor}
\usepackage{subfig}
\usepackage{comment}
\definecolor{mycolor1}{rgb}{0.00000,0.44700,0.74100}%
\definecolor{mycolor2}{rgb}{0.85000,0.32500,0.09800}%
\definecolor{mycolor3}{rgb}{0.92900,0.69400,0.12500}%
\definecolor{mycolor4}{rgb}{0.49400,0.18400,0.55600}%
\definecolor{mycolor5}{rgb}{0.46600,0.67400,0.18800}%
\definecolor{mycolor6}{rgb}{0.30100,0.74500,0.93300}%
\definecolor{mycolor7}{rgb}{0.63500,0.07800,0.18400}%
\definecolor{armygreen}{rgb}{0.29, 0.33, 0.13}

\definecolor{color_marks}{rgb}{0.85000,0.32500,0.09800}
\newcommand\numberthis{\addtocounter{equation}{1}\tag{\theequation}}
\newtheorem{rem}{Remark}
\begin{document}
%
\title{Identification and optimal control strategies\\for the transversal splitting of ultra-cold Bose gases 
}
%
%
%

\author{Nikolaus W\"urkner, Yevhenii Kuriatnikov, Karthikeyan Kumaran, M Venkat Ramana, J\"{o}rg Schmiedmayer, Andreas Kugi, Maximilian Pr\"ufer, Andreas Deutschmann-Olek\thanks{N. W\"urkner, A. Kugi, and A. Deutschmann-Olek are with the Automation and Control Institute, TU Wien, Gußhausstraße 27-29, 1040 Vienna, Austria. (e-mail:
\{wuerkner,kugi,deutschmann\}@acin.tuwien.ac.at)}\thanks{Y. Kuratnikov, K. Kumaran, M. Venkat Ramana, J. Schmiedmayer and M. Pr\"ufer are with the Vienna Center for Quantum Science and Technology, Atominsitut, TU Wien, Stadionallee 2, 1020 Vienna, Austria. (e-mail: maximilian.pruefer@tuwien.ac.at)}\thanks{A. Kugi is also with the AIT Austrian Institute of Technology, Giefinggasse 4, 1210, Vienna, Austria.}}

\markboth{ October 2025}%
{W\"urkner \MakeLowercase{\textit{et al.}}: arXiv}
%



\maketitle

\begin{abstract}
Splitting a Bose–Einstein condensate (BEC) is a key operation in fundamental physics experiments and emerging quantum technologies, where precise preparation of well-defined initial states requires fast yet coherent control of the condensate's nonlinear dynamics. This work formulates the BEC splitting process as an optimal feedforward control problem based on a physically interpretable, reduced-order model identified from limited experimental data. We introduce a systematic calibration strategy that combines optimal experiment selection and constrained nonlinear parameter estimation, enabling accurate system identification with minimal experimental overhead. Using this calibrated model, we compute energy-optimal trajectories via indirect optimal control to realize shortcuts to adiabaticity (STAs), achieving rapid transitions to the ground state of a double-well potential while suppressing excitations. Experiments confirm that the proposed control framework yields high-fidelity state transfers across multiple configurations, demonstrating its robustness and scalability for quantum control applications.
\end{abstract}


%
\IEEEpeerreviewmaketitle

\section{Introduction}

\IEEEPARstart{T}{he} application of control theory to quantum systems has given rise to the field of quantum control. 
Unlike their classical counterparts, quantum systems pose a number of unique control challenges. In particular, measuring a quantum system not only extracts information but also irreversibly disturbs the system's state. 
While weak measurements combined with feedback control schemes have been applied with striking success i.e., to stabilize photon-number states in a cavity~\cite{sayrin_real-time_2011}, implement quantum error correction schemes~\cite{campbell_series_2024}, and to prepare optically levitated particles in their quantum ground state~\cite{magrini_real-time_2021}. 
Additionally, many experimental measurement techniques are inherently destructive, rendering feedback control infeasible. 
As a result, (optimal) feedforward control strategies remain the dominant control approach. 
Over the past decade, the quantum control community has developed a number of tailored methods to obtain optimal feedforward trajectories, including gradient-descent techniques~\cite{khaneja_optimal_2005}, variants of Krotov's methods~\cite{reich_monotonically_2012} and (randomized) basis function approaches~\cite{doria_optimal_2011, caneva_chopped_2011,muller_one_2022}; recent developments are surveyed in~\cite{koch_quantum_2022}.
Despite substantial theoretical progress, experimental demonstrations of optimal control strategies remain comparatively scarce; notable examples include the generation of non-classical states~\cite{van_frank_interferometry_2014,larrouy_fast_2020,omran_generation_2019} and the implementation of logical qubits~\cite{heeres_implementing_2017}.
A key limitation arises from the fact that the performance of feedforward schemes crucially depends on the accuracy of the underlying system model. 
Yet realistic quantum systems are often described by complex, uncertain, and high-dimensional models that can exceed the computational capabilities of even the fastest classical computers. 
This motivates the exploration of alternative approaches, ranging from completely model-free online optimization methods~\cite{wigley_fast_2016}, to basis function approaches~\cite{doria_optimal_2011, caneva_chopped_2011}, and machine learning techniques to obtain data-driven models\cite{ma_machine_2025}.

Ultracold atoms have emerged as a pristine platform for exploring quantum effects in many-body systems.
Their long coherence times and precise experimental control make them ideal testbeds for studying fundamental quantum phenomena and for implementing quantum technologies.
A prerequisite for such experiments is the controlled preparation of well-defined quantum states, which naturally calls for the use of optimal control techniques. 

In this paper, we solve the problem of coherently splitting an elongated BEC into two with tailored optimal control strategies. 
Coherent splitting is a key step in many applications, as it allows the generation of two coherent condensates with interesting quantum properties~\cite{esteve_squeezing_2008,zhang_squeezing_2024}. 
This splitting ideally needs to balance two conflicting goals: On the one side, we want to split slowly in a quasi-adiabatic fashion, which avoids exciting the condensate and potentially produces highly entangled states. On the other side, we need to limit the ramp time as decoherence and loss in the BEC increase over time.
However, one can obtain fast yet adiabatic state transitions through carefully designed splitting trajectories~\cite{hohenester_optimizing_2009}, co-called shortcuts to adiabaticity~\cite{torrontegui_shortcuts_2012,campo_shortcuts_2012,guery-odelin_shortcuts_2019}.

A key contribution of this work is the development of an efficient identification approach that enables the accurate calibration of a reduced-order model for the transversal splitting process using limited experimental data. 
The resulting model serves to compute optimal transitions into split states of the condensate while introducing minimal energy.

This paper is organized as follows: Sec.~\ref{sec:mod_red} details the experimental system, its reduced-order model, and the available measurements in detail.
Sec.~\ref{sec:opt_exp} develops methods to optimally select a set of calibration experiments, limiting redundant information.
This selection is then used in Sec.~\ref{sec:calib} to calibrate the reduced-order model to measurements for two different configurations.
Building on the calibrated model, Sec.~\ref{sec:STA} presents a method for achieving optimal splitting of the BEC, demonstrating excellent performance in experiments across numerous scenarios, including ground-state to ground-state transitions as well as canceling known excitations.
A comprehensive overview of further experimental results and performance is explored in~\cite{kuriatnikov_fast_2025}, together with an evaluation of the impact on the quantum properties of the resulting state.

\begin{figure}
    \centering
    \includegraphics[width=1\linewidth]{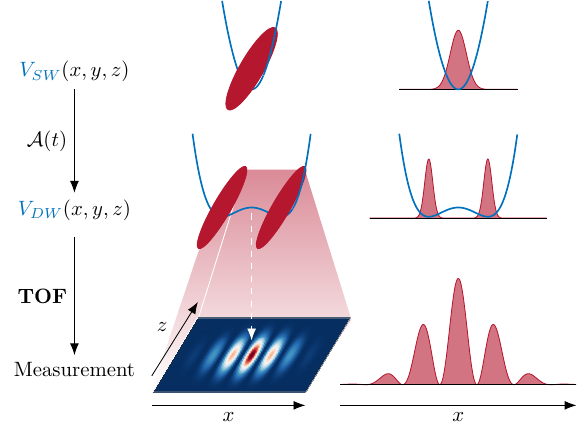}
    \caption{Schematic of the transversal splitting procedure run on the experimental setup. The left column shows the BEC being split by applying RF-dressing controlled via the normalized RF amplitude $\mathcal{A}(t)$, changing from a single-well potential $V_{SW}$ to a double-well potential $V_{DW}$. This potential is then deactivated; in the following time-of-flight phase, expansion of the BEC creates the measured two-dimensional interference pattern. By approximating the influence of the second transversal direction $y$ and averaging over the longitudinal direction $z$, we arrive at the reduced one-dimensional representation along the $x$-direction, shown in the right column.}
    \label{fig:procedure}
\end{figure}

\section{System description and mathematical model}
\label{sec:mod_red}
We consider a Bose-Einstein condensate (BEC) \cite{griffin1996bose} of ultra-cold Rubidium ($^{87}$Rb) atoms which is routinely prepared, manipulated and measured with high precision.
In our system, the trapping potential consists of a shallow confinement in the longitudinal ($z$-) direction and a tight confinement in the two transversal ($x$- and $y$-) directions.
The shape of the transversal potential can be adjusted using radio-frequency (RF) dressing~\cite{hofferberth_radiofrequency-dressed-state_2006,lesanovsky_adiabatic_2006, lesanovsky_manipulation_2006}; for example we can produce double-well potentials which allow for tunneling dynamics between the two wells.
We calibrate the RF dressing such that the trapping potential can be continuously transformed from a single into a double-well potential in $x$-direction with a single control parameter, the normalized RF amplitude $\mathcal{A}$. 
After splitting the condensate, the trapping potential is switched off and we measure the atomic density distribution after time-of-flight \cite{bucker_single-particle-sensitive_2009}; this experimental procedure is illustrated in the left column of Fig.~\ref{fig:procedure}.

In the following, we present a reduced-order model for the  transversal splitting process.
The dynamics along the $z$-direction are nearly two orders of magnitude slower than in the two transversal directions, therefore it is reasonable to assume the BEC remains stationary along $z$ during the splitting process.
Furthermore, the tightly confined $y$-direction is approximately decoupled and can be neglected when controlling the splitting process in $x$-direction. 
These assumptions justify a one-dimensional description of the BEC (illustrated in the right column of Fig.~\ref{fig:procedure}). 

The evolution of the transversal wavefunction $\Psi(x,t)$ of the condensate in the low temperature limit is governed by the Gross-Pitaevskii Equation (GPE), which is a non-linear Schrödinger-type equation. 
We write it in rate form (i.e., the right-hand side is expressed in units of angular-frequency) as
\begin{align}
    i\frac{\partial \Psi}{\partial t}= \Big(\Tilde{\Delta}_{xx} + V + g_{\perp}|\Psi|^2\Big)\Psi,\label{eq:GPE}
\end{align}
using the scaled Laplace operator
\begin{align*}
    \Tilde{\Delta}_{xx} = -\frac{\hbar}{2m}\frac{\partial^2}{\partial^2 x},
\end{align*}
the reduced Planck constant $\hbar$, the mass $m$, the transversal interaction strength $g_{\perp}$, and the effective transversal potential $V(x,\mathcal{A}(t))$ parametrized by the normalized RF amplitude $\mathcal{A}(t)\in[0,1]$, which serves as our control input. 
Note that numerical solutions of the one-dimensional GPE in \eqref{eq:GPE} and its corresponding optimal control problem are considerably faster to compute than in the three-dimensional case~\cite{mennemann_optimal_2015}.

Both the effective transversal potential $V(x,\mathcal{A})$ and the transversal interaction strength $g_\perp$ have to be determined for each experiment configuration.
The trapping potential can be calculated by simulating the magnetic field~\cite{lesanovsky_adiabatic_2006,lesanovsky_manipulation_2006}. 
Likewise, the transversal interaction strength $g_{\perp}$ depends on the atom density of the BEC and can be approximated from the true three-dimensional trapping configuration and the atom number~\cite{gerbier_quasi-1d_2004}.
However, these first-principle models are subject to inaccuracies from the underlying approximations and uncertainties.
Additionally, its relation with the transversal potential $V(x,\mathcal{A}(t))$ in the reduced one-dimensional description \eqref{eq:GPE} is not straight forward and the large number of physical parameters makes it very difficult to calibrate the potential to measurement data. 
As a result, even the most carefully constructed estimates of $g_\perp$ and $V(x,\mathcal{A})$ before identification, failed to provide the accuracy required for optimal feedforward control.

\subsection{Ansatz for $V(x,\mathcal{A})$}
\label{subsec:model}
To obtain a model for the effective transversal potential, we select a set of simple basis functions in position $x$ and normalized RF amplitude $\mathcal{A}$.
While high-order polynomials are often used, their coefficients can yield unphysical behavior of \eqref{eq:GPE} and result in non-confining potentials. 
Instead, we seek a parametrization that ensures physicality while maintaining interpretable parameters.

\begin{figure}[!t]
\centering
\includegraphics[width=\linewidth]{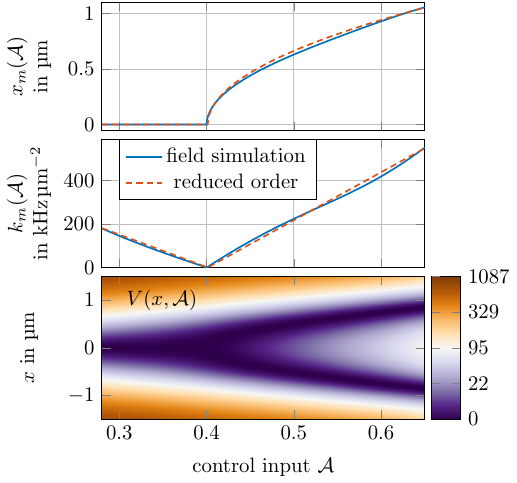}
\caption{Properties of the transversal potential obtained from first-principles-based field simulation (solid blue line) and the reduced-order model (dashed red line). The top panel shows the position of the minima $x_m(\mathcal{A})$ as a function of the normalized RF amplitude $\mathcal{A}$. The middle panel shows the curvature $k_m(\mathcal{A})=\frac{\partial^2}{\partial x^2}V\Big(x_m(\mathcal{A}),\mathcal{A}\Big)$. The bottom panel shows the resulting effective potential $V(x,\mathcal{A})$ along $x$.}
\label{fig_features}
\end{figure}

The simplest polynomial in space able to represent a symmetric double-well potential is an even quartic polynomial
\begin{align} \label{eq:effective_potential}
    V(x,\mathcal{A}) = a_4(\mathcal{A})x^4 + a_2(\mathcal{A})x^2,
\end{align}
where for positive $a_4(\mathcal{A})>0$, the sign of $a_2(\mathcal{A})$ determines if $V(x,\mathcal{A})$ is a single well ($\geq0$) or a double well ($<0$).
To choose an efficient ansatz for the coefficients $a_4(\mathcal{A})$ and $a_2(\mathcal{A})$,
one can investigate characteristic features of the simulated first-principles-based three-dimensional potential~\cite{lesanovsky_adiabatic_2006,lesanovsky_manipulation_2006}, depicted in Fig.~\ref{fig_features}.
Two specific features are the (positive) position of the local minima $x_m(\mathcal{A}) = \operatorname{arg min}_{x\geq 0} V(x,\mathcal{A})$ and the curvature $k_m(\mathcal{A})=\frac{\partial^2}{\partial x^2}V(x_m(\mathcal{A}),\mathcal{A})$ at these minima,
which can be functionally approximated as
\begin{subequations}\label{eq:potential_features}
    \begin{align}
        x_m(\mathcal{A})\approx 
        \begin{cases}
            0,              & \mathcal{A} \leq \mathcal{A}_s,\\
            c\sqrt{\mathcal{A}-\mathcal{A}_s},  & \mathcal{A} > \mathcal{A}_s
        \end{cases},
    \end{align}
    and
    \begin{align}
        k_m(\mathcal{A}) \approx 
        \begin{cases}
            \kappa_1(\mathcal{A}-\mathcal{A}_s), & \mathcal{A} \leq \mathcal{A}_s,\\
            \kappa_2(\mathcal{A}-\mathcal{A}_s), & \mathcal{A} > \mathcal{A}_s
        \end{cases},
    \end{align}
\end{subequations}
with the position scaling factor $c$, the splitting parameter $\mathcal{A}_s$ and the two slope values $\kappa_1<0$ and $\kappa_2>0$.
The coefficients $a_2(\mathcal{A})$ and $a_4(\mathcal{A})$ follow directly from these approximations in the form
\begin{subequations}
\label{eq:coeffs}
    \begin{align}
        a_2(\mathcal{A}) = 
        \begin{cases}
            \frac{\kappa_1}{2}(\mathcal{A}-\mathcal{A}_s), & \mathcal{A} \leq \mathcal{A}_s,\\
            -\frac{\kappa_2}{4}(\mathcal{A}-\mathcal{A}_s), & \mathcal{A} > \mathcal{A}_s
        \end{cases}\label{eq:a2}
    \end{align}
    and 
    \begin{align}
        a_4(\mathcal{A}) = \frac{\kappa_2}{8c^2}.\label{eq:a4}
    \end{align}
\end{subequations}
Here, $\mathcal{A}_s$ marks the value where $a_2(\mathcal{A})$ changes its sign, indicating the transition from single- to double-well potential.
\begin{rem} \label{rem:A_s_value}
There are two different parameter values that could be defined as the splitting parameter: The first is $\mathcal{A}_s$, where the potential starts displaying two wells, and the other is the parameter value at which the oscillation in the two split wells becomes dominant over the one in the confining single well.
\end{rem}
Thus, the ansatz for the effective transversal potential is defined by four coefficients $\kappa_1$, $\kappa_2$, $c$,and $\mathcal{A}_s$.
Fitting this ansatz to results from first-principles simulations according to \cite{lesanovsky_adiabatic_2006,lesanovsky_manipulation_2006} shows excellent agreement, as illustrated in Fig.~\ref{fig_features}.

\subsection{Dynamics of splitting} \label{subsec:dynamics_splitting}
\begin{figure}[!t]
\centering
\includegraphics[width=0.96\linewidth]{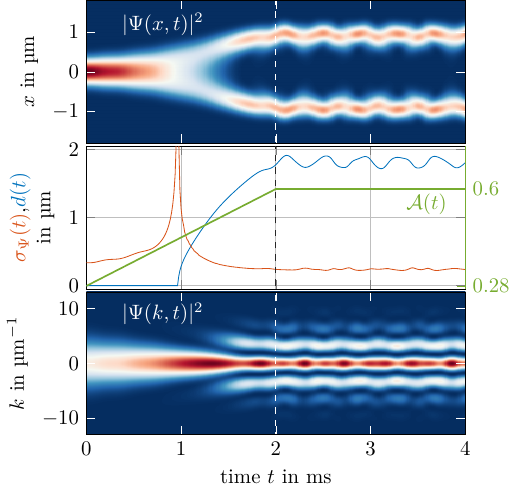}
\caption{Simulation results of a linear change from $\mathcal{A}(0)=0.28$ to $\mathcal{A}(2~\si{\milli\second})=0.6$ and the subsequent hold time using the modeled potential. The top panel shows the evolution of the transversal wavefunction density $|\Psi(x,t)|^2$ over time, the central panel illustrates the input trajectory $\mathcal{A}(t)$ as well as the fitted width $\sigma_\Psi(t)$ and inter-well distance $d(t)$ of the in-situ wavefunction. The bottom panel displays the corresponding transversal momentum of the BEC's mean field, calculated using the Fourier transformation of the transversal wavefunction $\Psi(x,t)$. Simulated using interpolated values of the first-principles model}
\label{fig_validation}
\end{figure}

A dynamic simulation of a splitting process is depicted in Fig.~\ref{fig_validation}, with the normalized RF amplitude $\mathcal{A}$ initially ramped up and subsequently held constant.
As $\mathcal{A}$ increases, the density $|\Psi(x,t)|^2$ of the BEC separates, in the subsequent static potential, each cloud oscillates within its individual well.
Although the splitting process excites higher-order modes, a useful analytic approximation is provided by fitting two Gaussian wave packets
\begin{align}
    \Psi(x) \approx A_\Psi\Bigg[\exp\Big(\frac{-(x-\frac{d}{2})^2}{2\sigma_\Psi^2}\Big)+\exp\Big(\frac{-(x+\frac{d}{2})^2}{2\sigma_\Psi^2}\Big)\Bigg],\label{eq:double_gauss}
\end{align}
with the in-situ width $\sigma_\Psi$ and the so-called inter-well distance $d$. 
Oscillations of $\sigma_\Psi$ and $d$ are typically referred to as \emph{breathing} and \emph{sloshing} in literature~\cite{dalfovo_theory_1999}.
The wave function can be analyzed in momentum space, accessible via the Fourier transform of $\Psi(x)$.
The corresponding momentum representation for~\eqref{eq:double_gauss} is
\begin{align}
    |\hat{\Psi}(k)|^2=2A_\Psi^2\sigma_\Psi^2\exp\big(-\sigma_\Psi^2k^2\big)\big(1+\cos(dk)\big),\label{eq:fit_gauss_cos}
\end{align}
with the wave number $k$.
Changes in the envelope width of the momentum profile reflect breathing, while displacement of its minima represents sloshing.

\subsection{Time-of-flight (TOF) measurements}
\label{subsec:TOF_measurement}
Following release from the trapping potential, the atom clouds expand for time $T_{TOF}$ and eventually overlap, which results in an interference pattern which is exemplified in Fig.~\ref{fig:procedure}.
In the idealized case where $g_\perp = 0$, the density distribution after infinite $T_{TOF}$ reflects the initial momentum distribution. 
For sufficiently long $T_{TOF}$ and high momenta, the momentum distribution of the atoms can therefore be measured to a good approximation. 
The wavenumber $k_{TOF}$ can then be calculated from the measurement position using
\begin{align}
    k_{TOF} = \frac{m}{\hbar T_{TOF}}x_{TOF}.\label{eq:balistic_k}
\end{align}
In principle, one could directly recover the momentum density $|\hat{\Psi}(k,t)|^2$ at any time $t$ by averaging repeated measurements.

\begin{figure}[!t]
\centering
\includegraphics[width=1\linewidth]{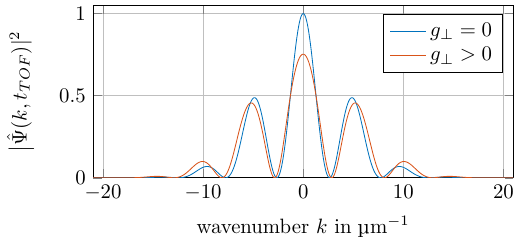}
\caption{Integrated momentum density for TOF with nonlinear interaction ($g_\perp>0$) and no interaction ($g_\perp=0$) compared for the ground-state wavefunction for $V(x,\mathcal{A}=0.5)$, normalized for illustrative purposes. 
The broadening of the envelope and shift of the minima to higher k resembles a BEC that is split less and confined more tightly.}
\label{fig_momentum_comparison}
\end{figure}

However, when $g_\perp>0$, nonlinear interactions during the initial stages of the TOF cannot be neglected.
These interactions lead to a broadening of the observed envelope and an outward displacement of the interference fringes, as illustrated in Fig.~\ref{fig_momentum_comparison} with simulations using \eqref{eq:GPE}, making the condensate appear less split and more tightly confined along $x$-direction.
Accurate modeling would require numerically intensive computation of the complete three-dimensional expansion to account for the additional dilution of the condensate, which is infeasible due to the high computational effort and would require accurately known potentials along the remaining spatial directions.
Fortunately, the relation between the idealized $d$ and its value after nonlinear expansion is a static, monotonous, nonlinear map. 
Thus, the frequency of sloshing oscillations is preserved in TOF measurements and can be leveraged for model identification purposes.

In summary, the combination of the one-dimensional GPE~\eqref{eq:GPE} and the approximated potential model \eqref{eq:coeffs} provides an accurate representation of the BEC transversal mean-field dynamics, fully characterized by five parameters $\mathbf{p}^T = [\kappa_1,\kappa_2,c,A_s,g_\perp]$.
For model-based feedforward control approaches, these parameters must be calibrated, measurements of the sloshing frequency can be used for to achieve this.

\section{Optimal selection of identification experiments}
\label{sec:opt_exp}

The model introduced in Sec.~\ref{subsec:model} has five parameters that need to be identified from measurement data. 
As discussed in Sec.~\ref{subsec:TOF_measurement}, directly extracting information from TOF measurements is limited to sloshing frequencies, which requires measurements at multiple times after the ramp. Furthermore, multiple experiment runs are needed at each time step to obtain high fidelity fits of~\eqref{eq:fit_gauss_cos}.
Since this process is very time-consuming, taking 2 to 3 hours for a single frequency depending on the desired accuracy, we need a systematic approach to select a set of experiments that yields sufficient information for model calibration while minimizing the total experimental and computational effort.

To excite measurable oscillations in the BEC for a particular potential configuration, we deploy three types of experiments:
In the first two, the BEC is split using the control input $\mathcal{A}$.
The specific input shapes are detailed in the top row of Fig.~\ref{fig_selected_exp}.
Both experiments induce sloshing oscillations, where the split condensates oscillate inside their respective wells, as elaborated in Sec.~\ref{subsec:dynamics_splitting}. 
Since these require the BEC to be split, they cannot probe the potential for $\mathcal{A} < \mathcal{A}_s$.
The third type excites the BEC through spatial displacements (offset) inside a chosen static potential with $\mathcal{A}=\mathcal{A}_T$, inducing an oscillation of the center of mass
\begin{align}
    \overline{x}(t) = \int x|\Psi(x,t)|^2\text{d}x,
\end{align}
enabling exploration of the potential for $\mathcal{A}<\mathcal{A}_s$.

\subsection{Measure of optimality}
To quantify the information yielded by a selection of experiments we derive a metric in the following.
For a specific experiment $\varepsilon_i$ and parameter vector $\mathbf{p}$, the measured frequency
$y_i$ is described as
\begin{align}\label{eq:output_equation_experiment}
    y_i = h_i(\mathbf{p},\varepsilon_i) + v_i,
\end{align}
where $h_i(\mathbf{p},\varepsilon_i)$ is a deterministic nonlinear mapping and $v_i$ is additive noise with covariance $q_i$ assumed uniform across measurements.
For a set of experiments $\mathcal{E}=\{\varepsilon_1,\dots,\varepsilon_{N_\mathcal{E}}\}\subset\mathcal{D}$, the vector of all measurements becomes 
\begin{align}\label{eq:output_equation_experiment_vec}
    \mathbf{y} = \mathbf{h}(\mathbf{p},\mathcal{E}) + \mathbf{v},
\end{align}
where $\mathbf{h}(\mathbf{p},\mathcal{E})$ is the vector-valued nonlinear mapping and $\mathbf{v}$ is additive measurement noise with covariance matrix $\mathbf{Q}$.
\begin{rem} \label{rem:multi_frequency}
Note that the output of a single experiment \eqref{eq:output_equation_experiment} can be directly extended to include multiple frequency components, which may be reasonable to account for excited higher-order spatial modes of the condensate. 
Indeed, this can be beneficial to capture weak double-well potentials in the vicinity of $\mathcal{A}_s$ more accurately, where oscillations of the original single-well and the newly formed double-wells coexist.
\end{rem}
Linearizing \eqref{eq:output_equation_experiment_vec} around $\mathbf{p}$ yields
\begin{align}
    \Delta\mathbf{y} = \mathbf{S}(\mathbf{p},\mathcal{E})\Delta\mathbf{p} + \mathbf{v},
\end{align}
with the sensitivity matrix
\begin{align}
    \mathbf{S}(\mathbf{p},\mathcal{E}) = \Big(\nabla_\mathbf{p}\mathbf{h}\Big)(\mathbf{p},\mathcal{E}).
\end{align}
The Fisher information matrix (FIM) 
for the linear system is
\begin{align}
    \mathcal{I}(\mathbf{p},\mathcal{E}) = (\mathbf{S}(\mathbf{p},\mathcal{E}))^T\mathbf{Q}\mathbf{S}(\mathbf{p},\mathcal{E}),
\end{align}
which provides a lower bound for the error covariance matrix of the parameter estimates~\cite{riley_mathematical_2006}.
Output scaling is used to mitigate effect the of varying value ranges~\cite{omlin_biogeochemical_2001}, normalizing the output via $\Tilde{\mathbf{y}} = \mathbf{Q}^{-\frac{1}{2}}\mathbf{y}$.
Aiming to remove the influence of different parameter value ranges, scaling the columns of $\mathbf{S}$ to unit norm yields the normalized FIM
\begin{align}
    \Tilde{\mathcal{I}}(\mathbf{p},\mathcal{E}) = (\Tilde{\mathbf{S}}(\mathbf{p},\mathcal{E}))^T\Tilde{\mathbf{S}}(\mathbf{p},\mathcal{E}).\label{eq:FIM_norm}
\end{align}
To select a set of optimal identification experiments, a scalar metric of the FIM such as A-, D-, or E-optimality~\cite{pukelsheim_optimal_2006} is generally used.
In particular, E-optimality balances the covariances of all parameters, ensuring robustness and good accuracy in each parameter estimate, by using the metric
\begin{align} \label{eq:e_optimality_metric}
    M_E(\mathbf{p},\mathcal{E}) = \frac{1}{\lambda_{\min}(\Tilde{\mathcal{I}}(\mathbf{p},\mathcal{E}))},
\end{align}
where $\lambda_{\min}(\cdot)$ denotes the minimum eigenvalue.
For the normalized FIM, \eqref{eq:e_optimality_metric} is directly related to the so-called \emph{collinearity index} given by $\rho_\mathcal{E} = \sqrt{M_e(\mathbf{p},\mathcal{E})}$, indicating the degree of parameter collinearity.
Values above 20 signal substantial linear dependence~\cite{brun_practical_2001}. 
Through the normalization introduced in~\eqref{eq:FIM_norm}, the collinearity index is naturally biased towards small number of experiments $N_\mathcal{E}$ to avoid linear dependence.
Hence, the optimal selection of identification experiments can then be formulated as the optimization problem
\begin{align}
    \mathcal{E^*} = \underset{\mathcal{E}\ \subset\ \mathcal{D}}{\arg\min}\ \rho_\mathcal{E},\label{eq:opt_GA}
\end{align}
which implicitly aims to keep $N_\mathcal{E^*} = |\mathcal{E}^*|$ as small as possible.

Each experiment $\varepsilon_i \in \mathcal{E}$ is classified by either one ($\mathcal{A}_T$) or two ($\mathcal{A}_T,T_R$) continuous parameter values depending on the chosen type.
To reduce the complexity of the optimization problem, these value spaces are discretized.
Still, exhaustively searching all combinations and calculating $\rho_{\mathcal{E}}$ is infeasible.
For example, one would need yield $2.7\cdot10^{46}$ unique combinations for $N_\mathcal{E} = 30$.
To numerically solve~\eqref{eq:opt_GA}, genetic algorithms (GAs) are particularly well suited as they can efficiently explore large, discrete search spaces and tolerates variable-length solution vectors~\cite{goldberg_genetic_1989,goldberg_investigation_1990}.

\subsection{Selection using genetic algorithm}
Our GA implementation combines variable-length chromosomes and random subset crossover.
The population consists of the members $\mathcal{E}_i, \ i=1,\dots,240$, each initialized as a random subset of 30 experiments.
Each GA iteration comprises:
\begin{enumerate}
    \item \textit{Fitness Evaluation}: Compute fitness, i.e., $\rho_{\mathcal{E}}$, for all members.
    \item \textit{Selection}: Remove the two-thirds with the highest $\rho_\mathcal{E}$.
    \item \textit{Crossover}: Randomly pair remaining members, merge their $\mathcal{E}$, and randomly select a subset with the subsets size chosen from a uniform distribution between parent set sizes.
    \item \textit{Mutation}: Swap each experiment in each member with a random other experiment with $0.1\%$ probability. Then, for each member, add ($0.5\%$) or remove ($0.6\%$) a random experiment.
\end{enumerate}
In every iteration, the member with the lowest $\rho_{\mathcal{E}}$ is stored, and the overall best is selected after the final iteration.
The rows of the sensitivity matrix (before normalization) can be precomputed using the nominal parameters, enabling the rapid evaluation of $\rho_\mathcal{E}$ $(4~\si{\milli\second})$ and permitting tens of thousands of iterations for robust results.

\subsection{Results}
\begin{figure}[!t]
\centering
\includegraphics[width=1\linewidth]{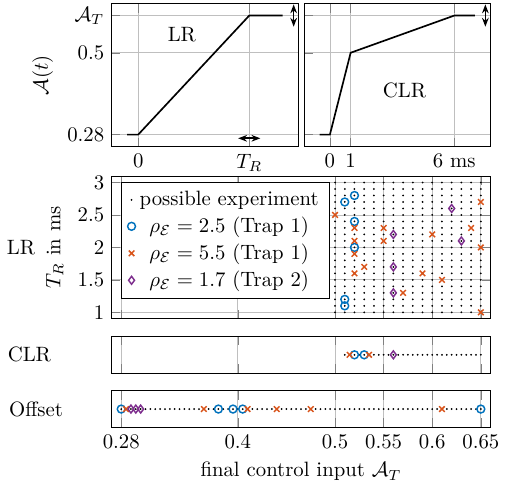}
\caption{The top two panels show the different types of experiments applied; linear ramps (LR) with the ramp time $T_R$ and value $\mathcal{A}_R$ varied in the left plot, and consecutive linear ramps (CLR) where only the end value $\mathcal{A}_R$ is varied in the right. Different combinations of selected experiments  are shown in the lower panels. The optimized selection for Trap~1 ($\rho_\mathcal{E}=2.5$) is marked \textcolor{mycolor1}{$\circ$}. The fittest member of the initial population ($\rho_\mathcal{E}=5.5$) is also shown, marked with \textcolor{mycolor2}{$\times$}. The optimized selection for Trap~2 ($\rho_\mathcal{E}=1.7$) is marked with \textcolor{mycolor4}{$\diamond$}. All selectable experiments are marked with gray dots.}
\label{fig_selected_exp}
\end{figure}

The optimal selection $\mathcal{E^*}$ for one GA run is shown in Fig.~\ref{fig_selected_exp}, achieving $\rho_\mathcal{E}=2.5$ with $N_{\mathcal{E^*}}=13$ compared to the fittest member of the initial population achieving  $\rho_\mathcal{E}=5.5$ with $N_{\mathcal{E}}=30$.
The resulting choice is also well motivated from a physical perspective: offset measurements mainly explore parameter ranges that remain inaccessible to sloshing measurements.
The splitting ramps explore higher parameter ranges and excite dynamics related to parameters not directly proportional to the sloshing frequency, such as $c$ and $g_{\perp}$.
For comparison, we also show the optimized result for a different experimental configuration (Trap~2), which features a larger splitting parameter $\mathcal{A}_s$.

\section{Calibration}
\label{sec:calib}

Given a chosen set of experiments $\mathcal{E} = \{\varepsilon_1,\dots,\varepsilon_{N_\mathcal{E}}\}$, we calibrate the simplified model introduced in Sec.~\ref{sec:mod_red} by minimizing the sum of squared frequency errors
\begin{align}
    J_y(\mathbf{p}) = \sum_{i=1}^{N_\mathcal{E}}\Big(h_i(\mathbf{p},\varepsilon_i)-y^m_i\Big)^2\label{eq:calib_cost},
\end{align}
with the measured frequencies $y^m_i$ of the specific experiment $\varepsilon_i$.
To ensure physical plausibility, we impose additional box constraints on the parameters $\mathbf{p}$.
Empirical estimates imply $-3000<\kappa_1<-500$, $500<\kappa_2<4000$, and $0.35<\mathcal{A}<0.45$.
From the first-principles model, we expect $1.5<c<3$ and a nominal value of $g_{\perp}=6.28$ for $4000$ trapped atoms, with an anticipated range of $4<g_\perp<10$.
These boundaries are collected in the vectors $\mathbf{p}_L$ and $\mathbf{p}_U$.

Hence, we obtain optimal parameters $\mathbf{p}^*$ by solving the constrained optimization problem
\begin{align*}
     (\mathbf{p}^*) = &\ \underset{(\mathbf{p})}{\arg\min}\ J_y(\mathbf{p})\numberthis\label{eq:opt_prob}\\ 
     & \text{s. t.}\\
     & \mathbf{p}_L \leq \mathbf{p} \leq \mathbf{p}_U
\end{align*}
using initial parameters $\mathbf{p}_0~=~[-1495,2210,2.115,0.4,6.28]^T$ for our specific setup.
We solve this problem numerically using an \textit{interior-point} algorithm.

\subsection{Calibration results on simulated data}
\begin{figure}[!t]
\centering
\includegraphics[width=1\linewidth]{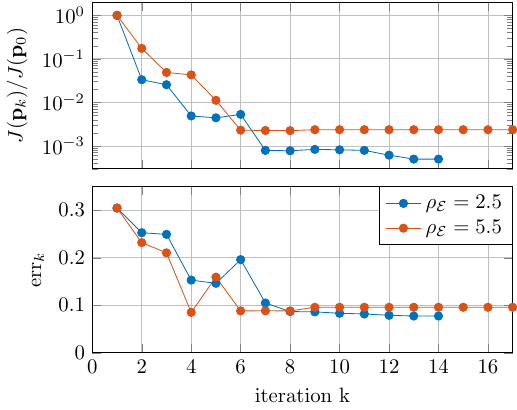}
\caption{Sum of squared relative errors $\text{err}_k$ and cost function $J_y(\mathbf{p_k})$ normalized to the initial cost, for the optimized selection of experiments with $\rho_\mathcal{E} = 2.5$ and the reference selections with $\rho_\mathcal{E} = 5.5$. Lower $\rho_\mathcal{E}$ ($2.5$) converges to lower values in the normalized cost and $\text{err}_k$ than higher $\rho_\mathcal{E}$ ($5.5$).}
\label{fig_calib_num}
\end{figure}

We first test our experiment selection on numerical data.
Using the simulation to generate the results of both selections of experiments in Fig.~\ref{fig_selected_exp} (i.e., optimal selection and best selection from initial population), and solve the nonlinear least squares problem \eqref{eq:opt_prob} for the simulated measurements.
Figure~\ref{fig_calib_num} displays the evolution of the cost function $J_y(\mathbf{p})$, alongside the normalized parameter error metric of the $k$-th iteration
\begin{align}
err_{k}=\sum_{i=1}^{5}\Big(\frac{p_{k,i}-p_i^*}{p_i^*}\Big)^2,
\end{align}
with the $i$-th element $p_{k,i}$ of the parameter vector and the $i$-th element $p_i^*$ of the true parameter vector.
As one can see, the optimal set of experiments 
achieves a greater reduction in $J_y(\mathbf{p})$ and converges closer to the correct parameters, while requiring less measurements.
Still, the set with a higher $\rho_\mathcal{E}$ also converges close to the target, indicating that reasonable calibration can be achieved even with a non-optimal experimental selection.

\subsection{Calibration results on experimental data}
\begin{figure}[!t]
\centering
\includegraphics[width=1\linewidth]{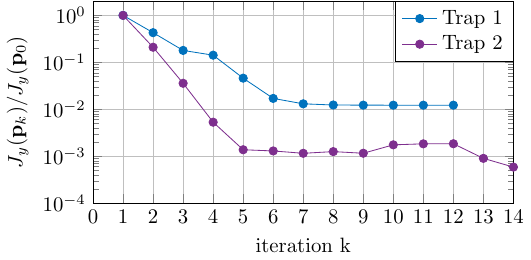}
\caption{Normalized cost function $J_y(\mathbf{p_k})$ over the steps of the calibration algorithm for the optimally-selected set of experiments for Trap~1 and Trap~2 in Fig.~\ref{fig_selected_exp} using measurement data.}
\label{fig_calib_cost_exp}
\end{figure}

\begin{figure}[!t]
\centering
\includegraphics[width=1\linewidth]{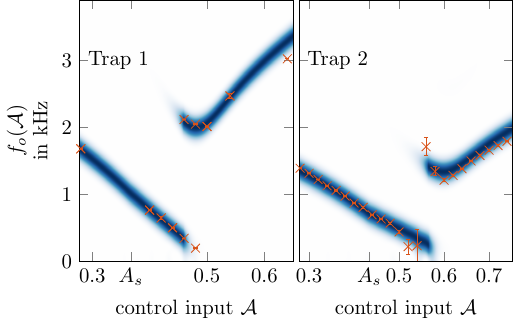}
\caption{Simulated frequency spectra for offset-type experiments with different control input values $\mathcal{A}$ compared to measurements of a validation data set showing the dominant frequencies of Trap~1 and Trap~2 as red crosses with error bars. For $\mathcal{A} > \mathcal{A}_s$, the dynamics transition from single-well dominant to double-well dominant.}
\label{fig_calib_offset_exp}
\end{figure}

\begin{figure}[!t]
\centering
\includegraphics[width=1\linewidth]{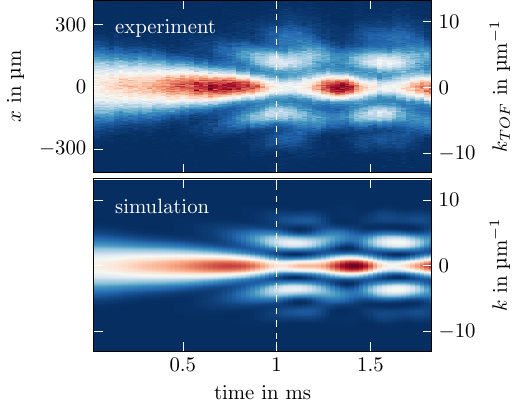}
\caption{The top panel shows the TOF measurements carpet from the experiment for a linear ramp from $\mathcal{A}(0)=0.28$ to  $\mathcal{A}(1~\si{\milli\second})=0.5$, using the relation in \eqref{eq:balistic_k} to convert the measured position $x$ to the wavenumber $k_{TOF}$. The bottom plot depicts the momentum density from the simulation for the same linear ramp. The simulation accurately reproduces the overall behavior; a small phase shift between the measurement and simulation is present. Additional discrepancies can be attributed to the nonlinear broadening during TOF not captured by the model.
}
\label{fig_calib_result_lin_1ms}%
\end{figure}

Applying the same calibration algorithm to experimental data yields the cost function evolution shown in Fig.~\ref{fig_calib_cost_exp}. 
The approach was tested on two experimental configuration (Trap~1 and Trap~2) to ensure that generality of the proposed method. Trap~2 is characterized by a larger splitting parameter $\mathcal{A}_s$ and lower values of $\kappa_1$, $\kappa_2$, and $c$ compared to Trap~1.
For verification purposes, a large number of experiments of the offset-type were performed for different values of the control parameter $\mathcal{A}$ and compared to simulated frequency spectra using the calibrated model, see Fig.~\ref{fig_calib_offset_exp}. 
The results indicate that the presented approach is effective for both experimental configurations. However, the accuracy of the model tend to degrade for higher values of the control parameter $\mathcal{A}$, with the model consistently predicting higher oscillation frequencies. We attribute this to the increasing effect of higher-order polynomial terms beyond the simple spatial ansatz \eqref{eq:effective_potential}.
Figure~\ref{fig_calib_offset_exp} further illustrates the presence of two significant oscillation frequencies in the intermediate regime where the dynamics transition from a dominant single-well into a dominant double-well potential. If this is the case for a given set of identification experiments $\mathcal{E}$, see Sec.~\ref{sec:opt_exp}, it is beneficial to include both dominant oscillation frequencies into the nonlinear least squares formulation \eqref{eq:calib_cost}, see also Remark~\ref{rem:multi_frequency}.
Figure~\ref{fig_calib_result_lin_1ms} compares the calibrated model to experimental measurements for a linear ramp with ramp time $T_R = 1~\si{\milli\second}$. The calibrated model captures the overall behavior very well except for a minor phase shift in the oscillations and nonlinear distortions due to TOF measurements.

\section{Optimal state transitions}
\label{sec:STA}
To achieve the desired coherent splitting of the BEC, it is necessary to design the control trajectory $\mathcal{A}(t)$ such that excitations of the condensate are minimized at the end of the process.
With the calibrated model developed in the preceding sections, we can implement model-based feedforward control schemes for this purpose.

\subsection{Optimal feedforward control}
A common strategy is to use an indirect optimization approach~\cite{winckel_computational_2008}.
Transferring the BEC from an arbitrary initial state $\Psi_0(x)$ into the ground state of a final potential $V(x,\mathcal{A}_T)$ is equivalent to steering the system \eqref{eq:GPE} into a state that minimizes the corresponding energy 
\begin{align*}
    E(\Psi,\mathcal{A}_T) = \int_{-\infty}^{\infty}\Psi^*\Big(\Tilde{\Delta}_{xx} + V(x,\mathcal{A}_T) + \frac{g_{\perp}}{2}|\Psi|^2\Big)\Psi\text{d}x.
\end{align*}
Because infinitely many trajectories could achieve this transition, a regularization term with weight $\gamma$ is typically introduced to penalize fast fluctuations of the control input $\mathcal{A}(t)$, yielding the cost function
\begin{align}
    J\big(\mathcal{A}(t)\big) = E(\Psi(x,T),\mathcal{A}(T)) + \frac{\gamma}{2}\int_{0}^{T}\Big(\frac{\partial \mathcal{A}(t)}{\partial t}\Big)^2\text{d}t.
\end{align}
The optimal control problem is thus defined as
\begin{align}\label{eq:OC_problem}
    \mathcal{A}^*(t) = &\underset{\mathbf{\mathcal{A}}(t)}{\arg\min}\ J(\mathcal{A}(t))
\end{align}
subject to the GPE~\eqref{eq:GPE} and the boundary conditions
\begin{subequations}\label{eq:boundary_opt}
\begin{align}
    \mathcal{A}(0) &= \mathcal{A}_0\\
    \mathcal{A}(T) &= \mathcal{A}_T\\
    \Psi(x,0) &= \Psi_0(x).
\end{align}
\end{subequations}
Using variational calculus, one obtains necessary first-order optimality conditions (see, e.g., \cite{hohenester_optimal_2007,deutschmann-olek_optimal_2023} for details) with the adjoint state $p(x,t)$ in the form of a boundary value problem
\begin{subequations}\label{eq:bound_val}
\begin{align}
    i\frac{\partial \Psi}{\partial t}=& \Big(\Tilde{\Delta}_{xx} + V + g_{\perp}|\Psi|^2\Big)\Psi\\
    i\frac{\partial p}{\partial t}=& \Big(\Tilde{\Delta}_{xx} + V + 2g_{\perp}|\Psi|^2\Big)p + g_{\perp}\Psi^2p^*\\
    \gamma\frac{\partial^2\mathcal{A}}{\partial t^2}=&-\text{Re}\int_{-\infty}^{\infty}p^*\frac{\partial V}{\partial \mathcal{A}}\Psi\text{d}x\\
    ip(x,T) =& 2\Big(\Tilde{\Delta}_{xx} + V(x,\mathcal{A}_T) + g_{\perp}|\Psi(x,T)|^2\Big)\Psi(x,T),
\end{align}
\end{subequations}
subject to the conditions~\eqref{eq:boundary_opt}.
Choosing a $H^1$ inner product space \cite{winckel_computational_2008}, the gradient $\nabla J_\mathcal{A}(t)$ of the cost function with respect to $\mathcal{A}(t)$ is determined by the Poisson equation
\begin{align*}
    \frac{\partial^2 \nabla J_\mathcal{A}}{\partial t^2}  &= \gamma\frac{\partial^2\mathcal{A}}{\partial t^2}+\text{Re}\int_{-\infty}^{\infty}p^*\frac{\partial V}{\partial \mathcal{A}}\Psi\text{d}x\numberthis\label{eq:poisson_eq}\\
    \nabla J_\mathcal{A}(0) &= \nabla J_\mathcal{A}(T) = 0.
\end{align*}
Hence, the optimal control problem \eqref{eq:OC_problem} can be solved in an iterative fashion using gradient-descent or quasi-Newton methods.
For a specified initial state $\Psi_0(x)$ and an initial control input trajectory $\mathcal{A}_{init}(t)$ (which determines the boundary values $\mathcal{A}_0$, $\mathcal{A}_T$, and the ramp time $T$), the state $\Psi(x,t)$ is first obtained via forward solution of the GPE. Subsequently integrating the adjoint state equation backward in time for $p(x,t)$, one can compute the gradient from \eqref{eq:poisson_eq}.
Specifically, we use the limited memory BFGS algorithm~\cite{nocedal_updating_1980} to reduce the amount of storage needed as well as utilizing a more local approximation.

\subsection{Fast transversal splitting}

\begin{figure}[!t]
\centering
\includegraphics[width=1\linewidth]{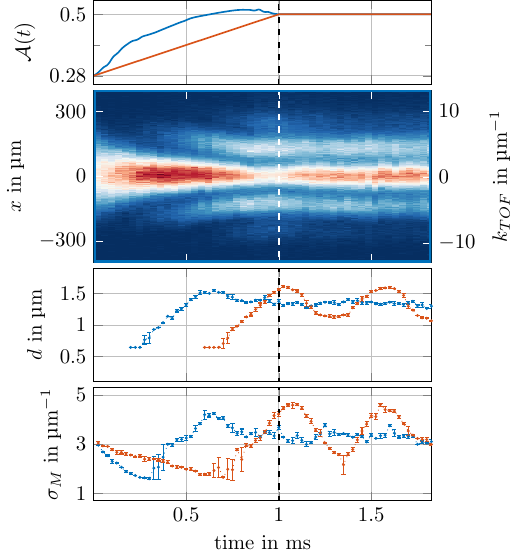}
\caption{Results of the STA solution designed with ramp time $T = 1~\si{\milli\second}$ from $\mathcal{A}_0 = 0.28$ to $\mathcal{A}_{T}=0.5$. The first panel highlights the differences between STA solution and the linear ramp. The TOF measurement carpet in the second panel shows that the STA solution moves the BEC into an almost steady state. The TOF measurement carpet for the linear ramp is presented in Fig.~\ref{fig_calib_result_lin_1ms}. The bottom two panels show a comparison of the fitted inter-well distance $d(t)$ and momentum width $\sigma_M(t)$ values for the linear ramp and the STA solution, illustrating a strong reduction of the excitations. To fit~\eqref{eq:fit_gauss_cos} to the measurement the measured positions where converted into the wavenumber space using~\eqref{eq:balistic_k}. }
\label{fig_result_oct_1ms}%
\end{figure}

Applying this algorithm to the scenario in Fig.~\ref{fig_calib_result_lin_1ms} yields the trajectory and experimental data shown in Fig.~\ref{fig_result_oct_1ms}.
The optimized trajectory differs from the linear ramp in an intuitive fashion. It approaches a double-well configuration sooner and temporally overshoots $\mathcal{A}_T$ to remove the excess excitations in the double-well.
In simulation, the optimal trajectory reduces the residual energy over ground-state energy by a factor of $407.07$ compared to the linear ramp.
Experimentally, this trajectory prepares a state with minimal density variations visible during the subsequent holding phase.
Such a protocol is termed a shortcut-to-adiabaticity (STA) \cite{guery-odelin_shortcuts_2019}, as it reproduces the effect of infinitely slow splitting but at much faster timescales.

\begin{figure}[!t]
\centering
\includegraphics[width=1\linewidth]{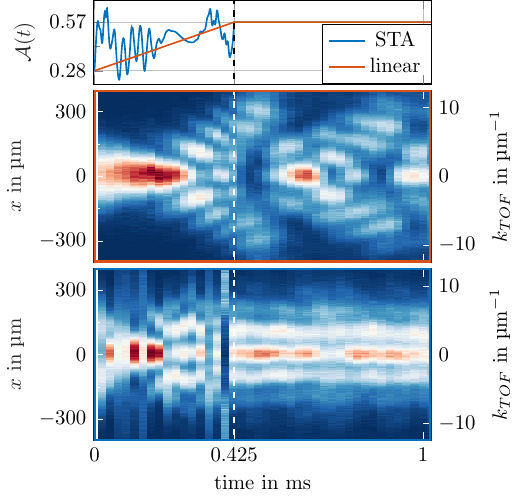}
\caption{Resulting TOF measurement carpets for the linear ramp and the STA solution with time $T = 0.425~\si{\milli\second}$ from $\mathcal{A}_0 = 0.28$ to $\mathcal{A}_{T}=0.57$. The oscillations excited by the linear ramp are so strong, that the BEC merges again periodically during the subsequent holding phase. Applying the STA solution, on the other hand, reduces the excitations so far as to reach a state that is very close to stationary.}
\label{fig_result_oct_0p425ms}%
\end{figure}

The algorithm further enables the engineering of trajectories that achieve splitting within time frames unattainable by simple linear ramps.
For instance, Fig.~\ref{fig_result_oct_0p425ms} shows results for a rapid transition from $\mathcal{A}_0 = 0.28$ to $\mathcal{A}_T = 0.57$ in $0.425~\si{\milli\second}$.
Here, a linear ramp produces a fractured BEC that is not split into two distinct clouds and crosses over the center of the potential during the holding phase, while the optimized trajectory successfully achieves splitting, closely resembling a STA solution, validating the predictive power of the model.

\begin{figure}[!t]
\centering
\includegraphics[width=1\linewidth]{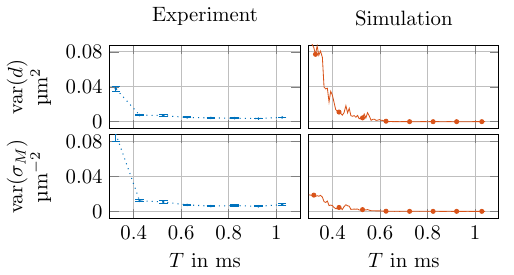}
\caption{Variances of the inter-well distance $d(t)$ and the Gaussian width $\sigma_M(t)$ in the holding phase after an optimized trajectory with different ramp times, comparing experiment and calibrated simulation model. The variances are reduced to near constant levels for ramp times over $0.7~\si{\milli\second}$ and start rising slightly for faster ramp times until blowing up for ramp times under $0.4~\si{\milli\second}$; showing something akin to a speed limit of the STA splitting. Note that a quantitative comparison of experiment and simulation is difficult since \eqref{eq:balistic_k} was used again to approximately convert the measured position to $k_{TOF}$ before fitting~\eqref{eq:fit_gauss_cos} and then calculating the variances.}
\label{fig_speed_limit}%
\end{figure}

Optimizing for different scenarios also allows exploration of the system's fundamental speed limits for state transition.
Fig.~\ref{fig_speed_limit} presents measured and simulated variances of the inter-well distance $d(t)$ and envelope width $\sigma_M(t)=(\sqrt{2}\sigma_\Psi(t))^{-1}$ during the post-ramp hold phase.
These variances represent the residual excitations present in the condensate. In simulation, these correspond directly to the added energy beyond the energy of the ground state, while in experimental data there are further contributions such as measurement noise that introduce an offset.
The variances remain at low and almost constant levels for ramp durations larger than $0.7~\si{\milli\second}$ and only raise slightly until $0.4~\si{\milli\second}$, below which they raise drastically and no suitable solution is found.
This suggests a minimum control time, characteristic for systems governed by wave or transport equations~\cite{meurer_control_2012}, due to the finite speed of expansion in the double-well potential that is directly related to its bounded curvature at the center.

\subsection{Canceling known excitations}
\begin{figure}[!t]
\centering
\includegraphics[width=1\linewidth]{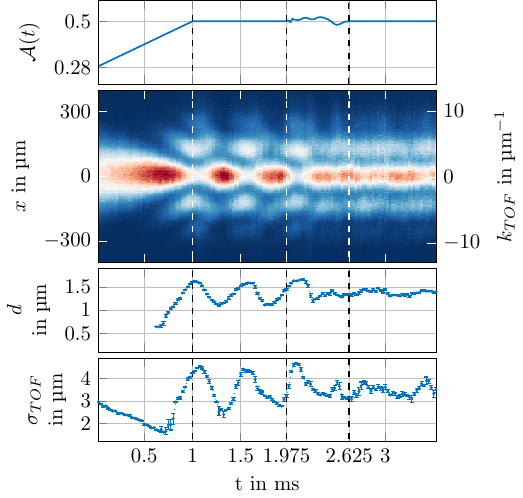}
\caption{Using optimal control sequences to cancel know excitations. The BEC is split and oscillations are excited by a linear ramp from $\mathcal{A}_0=0.28$ to $\mathcal{A}(1~\si{\milli\second}) = 0.5$; this potential is then held until $t=1.975~\si{\milli\second}$, at which point an optimized input trajectory of duration $T=0.65~\si{\milli\second}$ is applied. Comparing the experimental results before and after clearly shows that the BEC transitions from a highly excited to an almost stationary state. Note that the optimized sequence was designed for a nominal hold time of $1~\si{\milli\second}$, see Fig.~\ref{fig_hold_time_damping}.}
\label{fig_result_damping}%
\end{figure}

The optimal feedforward control algorithm can be further used to remove known (deterministic) excitations present in the condensate.
To illustrate this, we excite the condensate with a linear ramp from $\mathcal{A}_0=0.28$ to $\mathcal{A}(1~\si{\milli\second})=0.5$ and hold the potential constant for another millisecond. Using the calibrated model, we can predict the state at $t=2~\si{\milli\second}$. Starting from this excited initial state, we calculate an optimal control signal that steers the condensate back into its ground state. The complete sequence and experimental results are illustrated in Fig.~\ref{fig_result_damping}, confirming that existing oscillations are suppressed to negligible levels.

\begin{figure}[!t]
\centering
\includegraphics[width=1\linewidth]{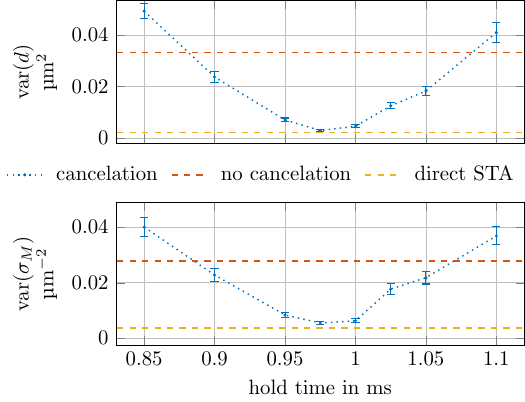}
\caption{The impact of timing on the cancellation protocol. The blue dotted line shows the residual variances of the optimal feedforward, designed for a nominal hold time of $1~\si{\milli\second}$, when applied using different actual hold times. For reference, the variance achievable through a direct STA to $\mathcal{A}(1~\si{\milli\second})=0.5$ is shown as a yellow dashed line and the variance from the linear ramp without cancellation is shown as a red dashed line. As one can see, the cancellation protocol achieves approximately the same performance as direct STAs. Notably, the minimum value of residual excitations is obtained by reducing the hold time by 0.025~\si{\milli\second} compared to the nominal value, which can be traced back to the mismatch between experiment and simulation, particularly in the phase shift shown in Fig.~\ref{fig_calib_result_lin_1ms}.
}
\label{fig_hold_time_damping}%
\end{figure}

Due to a residual model-plant mismatch, in particular the phase mismatch visible in Fig.~\ref{fig_calib_result_lin_1ms}, the predicted state at $t=2~\si{\milli\second}$ will not be perfectly accurate.
This leads to the trajectory designed for a hold time of $t=1~\si{\milli\second}$ performing best at slightly shorter hold times, see Fig.~\ref{fig_hold_time_damping}, highlighting how sensitive the performance is to the exact timing of the control signals.
Oscillations are suppressed significantly in the vicinity of the designed hold time of $t=1~\si{\milli\second}$, with a minimal residual excitation at hold time of $0.975~\si{\milli\second}$, which is also the hold time using in Fig.~\ref{fig_result_damping}.
For large deviations of the hold time, the optimized trajectory may inadvertently enhance oscillations.
Note that the remaining excitations of this procedure are comparable to direct STA trajectories, highlighting the efficacy of the calibrated model for control purposes.

\section{Conclusion}
In this paper, we introduced a reduced-order model for the adjustable trapping potential employed to split Bose-Einstein Condensates (BECs).
By capturing essential features of the potential using a simple, physically motivated ansatz, the model enables efficient model-based control to optimize transitions into desired ground states.
Our reduced-order model demonstrates good agreement with the simulations and uses parameters that are easy to constrain, ensuring that the resulting potentials remain physical.

To accurately calibrate these parameters, we propose a targeted experimental design strategy that works within the limitations imposed by the nonlinear interactions during time-of-flight.
Utilizing a genetic algorithm, we identified informative sets of experiments that yield sufficient data for calibration while minimizing experimental overhead.
This approach enables rapid adaptation of the model to varying experiment configurations, as demonstrated by results from two distinct scenarios.

Finally, we developed a optimal feedforward controls using indirect optimization, leveraging the calibrated model to achieve state transitions from an arbitrary known state to a desired ground state, including protocols that represent so-called shortcuts-to-adiabaticity . The resulting optimal control-driven splitting protocol provides a flexible and robust method for initializing low-entropy, entangled states, which are crucial for quantum-enhanced metrology and serve as fundamental building blocks for many-body quantum simulations.

The next steps will be to exert full control over the quantum properties of the splitting process, including quantum fluctuations. 
First calculations for zero dimensional systems in \cite{Grond_OCTnumberSqueezing2009} show that number squeezing can be achieved much faster then in adiabatic problems. This was verified for global observables in \cite{kuriatnikov_fast_2025}.
By further including excitations along the longitudinal direction, the splitting process moves through the so-called strongly correlated regime of the Sine-Gordon quantum field theory (SG-QFT) \cite{cuevas2014sine,gritsev2007linear,schweigler2017experimental}, which is beyond what can be calculated on classical computers. 
Consequently, preparing desired initial states of the SG-QFT will require to extract local, approximate models from measurement data that can be employed a closed-loop optimization schemes. 
\section*{Acknowledgment}
This research was funded by the Austrian Science Fund (FWF) through the Cluster of Excellence \emph{Quantum Science Austria} (quantA) [10.55776/COE1] and the principal investigator project ConQF [10.55776/P36236] and the QUANTERA project MENTA
(FWF: I-6006). M.P. has received funding from Austrian Science Fund (FWF) [10.55776/ESP396; QuOntM].

\ifCLASSOPTIONcaptionsoff
  \newpage
\fi



%
\bibliography{references}
\bibliographystyle{IEEEtran}

%








\end{document}